\documentclass[aps,prl,twocolumn,showpacs,groupedaddress]{revtex4}

\usepackage{graphicx}
\usepackage{dcolumn}
\usepackage{bm}

\begin{document}


\title{ 
Fluid Vesicles with Viscous Membranes in Shear Flow}

\author{Hiroshi Noguchi}
\author{Gerhard Gompper}
\affiliation{
Institut f\"ur Festk\"orperforschung, Forschungszentrum J\"ulich, 
52425 J\"ulich, Germany}


\begin{abstract}
The effect of membrane viscosity on the dynamics of vesicles in  
shear flow is studied. We present a new simulation technique, which combines 
three-dimensional multi-particle 
collision dynamics for the solvent with a dynamically-triangulated
membrane model. Vesicles are found to transit 
from steady tank-treading to unsteady tumbling motion with increasing
membrane viscosity. Depending on the reduced volume and membrane viscosity,
shear can induce both discocyte-to-prolate and prolate-to-discocyte 
transformations.
This dynamical behavior can be understood from a simplified model.
\end{abstract}


\maketitle

The dynamical behavior of vesicles --- closed lipid membranes in aqueous 
solution --- under shear flow 
is an important subject not only of fundamental research but also in 
medical applications.  For example, in microcirculation,
the deformation of red blood cells reduces the flow resistance of microvessels.
In diseases such as diabetes mellitus and sickle cell anemia,
red blood cells have reduced deformability and often block microvascular flow.  
Although red blood cells do not have a nucleus and other intracellular 
organelles,
they are of more complex than simple lipid vesicles, since their plasma membrane
has an attached spectrin network, which modifies its elastic and rheological
properties. 
The deformability of cells and vesicles is determined by their shape, 
the viscosity of the internal fluid, and the viscoelasticity of the 
membrane \cite{chie87}.

The dynamical behavior of vesicles in shear flow has been studied experimentally
\cite{chie87,fisc78}, theoretically \cite{kell82,tran84}, and numerically 
\cite{krau96,pozr03,beau04}.
The vesicle shape is determined by the competition of the curvature elasticity 
of the membrane, the constraints of constant volume $V$ and constant 
surface area $S$, and the external hydrodynamic forces.
One of the difficulties in theoretical studies of the hydrodynamic effects 
on the vesicle dynamics is the boundary condition for the embedding
fluid on the vesicle surface, which changes its shape dynamically.
In some previous studies, a fluid vesicle was therefore modeled as an 
ellipsoid with fixed shape \cite{kell82}. More recently, the 
time-evolution of the shape was studied numerically using a boundary 
integral method in 
three spatial dimensions \cite{krau96} or an advected-field 
method in two spatial dimensions \cite{beau04}. 
The red blood cell membrane has also been modeled as an elastic capsule of 
discocyte shape \cite{pozr03}.  

Two types of dynamics have been found in these studies, a steady state 
with a tank-treading 
motion of the membrane and a finite inclination angle with the flow 
direction, and an unsteady state with a tumbling (flipping) motion.
A transition 
from tank-treading to tumbling motion with an increasing viscosity of the 
internal fluid has been predicted 
for fluid vesicles with fixed ellipsoidal shape in three dimensions 
\cite{kell82}, and with the advected-field method in two dimensions 
\cite{beau04}.
When the shape is relaxed dynamically in three dimensions, all discocyte 
vesicles were 
surprisingly found to transform into prolates in shear flow, 
even for the smallest shear rates studied ~\cite{krau96}.

In this letter, we focus on the effect of the membrane viscosity on the  
dynamics of vesicles in shear flow. This is an important question, because
the membrane of red blood cells, for example, becomes more viscous on aging 
\cite{tran84,nash83} or in diabetes mellitus \cite{tsuk01}. Experiments
indicate that the energy dissipation in the membrane is larger than that 
inside a red blood cell \cite{tran84}. Furthermore, it has been shown recently
that vesicles can not only be made from lipid bilayers, but also from bilayers
of block copolymers \cite{disc99}. These ``polymersomes'' have been shown
to have a membrane viscosity which is several orders of magnitude larger than
for liposomes \cite{dimo02}.

Several mesoscopic simulation techniques for fluid flow have been developed
in recent years. We present here the first simulation studies for a 
combination of mesoscopic model for
the solvent and a coarse-grained, dynamically-triangulated surface model 
for the membrane.
This approach has four main advantages: (i) The membrane is described 
explicitly, so that their properties like the viscosity can be varied
easily; (ii) thermal fluctuation of both the solvent and the membrane are
fully and consistently taken into account; (iii) the method can easily 
be generalized to more complex flow geometries; and (iv) no numerical
instabilities can occur.

We employ a particle-based hydrodynamics method \cite{male99,male00b,ihle01,kiku03,kiku03a,lamu01,ripo04} to simulate the solvent, 
which is called multi-particle collision dynamics (MPCD) 
\cite{lamu01,ripo04} or 
stochastic rotation dynamics \cite{ihle01,kiku03}. 
This method was applied, for example, to flow around a solid object 
\cite{lamu01} and to polymer dynamics \cite{male00b,ripo04}.
The fluids in the interior and exterior of the vesicle 
are taken to be the same, in particular to have the same  
viscosity $\eta_0$.

As the MPCD model is described in detail in 
Refs.~\cite{male99,male00b,ihle01,kiku03}, we can be very brief in 
explaining the mesoscopic simulation technique.
The solvent is described by $N_{\text s}$ point-like particles of mass 
$m_{\text s}$ moving in a rectangular box  of size $L_x\times L_y \times L_z$.
The algorithm consists of alternating streaming and collision steps. 
In the streaming step, the particles move ballistically and the position of 
each particle ${\bf r}_{i}$ is updated according to
\begin{equation}
{\bf r}_{i}(t+h)= {\bf r}_{i}(t) + {\bf v}_{i}(t) h,
\end{equation}
where ${\bf v}_{i}$ is the velocity of particle $i$ and $h$ 
is the time interval between collisions. In the collision step, the particles 
are sorted into cubic cells of lattice constant $a$.
The collision step consists of a stochastic rotation of the relative velocities 
of each particle in a cell, 
\begin{equation}
{\bf v}_{i}(t)= {\bf v}_{\text {cm}}(t) + 
           {\bf \Omega}(\varphi)({\bf v}_{i}(t)-{\bf v}_{\text {cm}}(t)),
\end{equation}
where ${\bf v}_{\text {cm}}$ is the velocity of the center of mass of all 
particles in the cell. The matrix ${\bf \Omega}(\varphi)$ rotates velocities by a 
fixed angle $\varphi$ around an axis, which is chosen randomly for each cell. 
In our simulation, the angle $\varphi=\pi/2$ is employed. We apply a 
random-shift procedure \cite{ihle01} before each collision step to ensure  
Galilean invariance.

For the membrane, we employ a dynamically-triangulated surface model 
\cite{gg:gomp97f}, in which the membrane is described by 
$N_{\text {mb}}$ vertices which are connected by tethers to form a 
triangular network. The vertices have excluded volume and mass 
$m_{\text {mb}}$. The shapes and fluctuations of the membrane are 
controlled by curvature elasticity with the energy \cite{helf73}.
\begin{equation}
H_{\text {cv}} = \frac{\kappa}{2} \int (C_1+C_2 )^2 dS,
\end{equation}
where $\kappa$ is the bending modulus, and $C_1$ and $C_2$ are the principal 
curvatures at each point of the membrane. The curvature energy is discretized
as described in Ref.~\cite{gomp96}. To model the fluidity of the 
membrane, tethers can be flipped between the two possible diagonals of two
adjacent triangles. These bond flips provide also a convenient way to vary
the membrane viscosity $\eta_{\text {mb}}$, because it increases 
with decreasing bond-flip rate. In contrast to previous studies of 
dynamically triangulated surfaces, which were all done by Monte Carlo 
simulations, we introduce a smooth bond-interaction potential, which 
makes the model amenable for molecular dynamics simulations. 

The solvent particles interact with the membrane in two ways. First, the
membrane vertices are included in the MPCD collision procedure \cite{male00b}.
Second, the solvent particles are scattered elastically or via bounce-back  
from membrane triangles. We use here the procedure suggested in  
Ref.~\cite{kiku03a} for a spherical particle.

\begin{figure}
\includegraphics{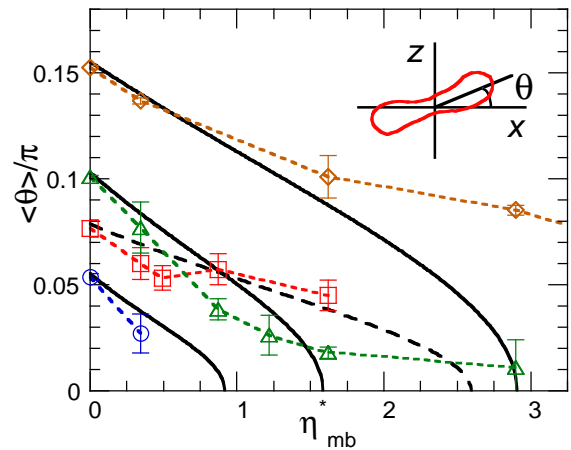}
\caption{  \label{fig:ang}
(Color online)
Dependence of the average inclination angle $\langle\theta\rangle$ 
($-\pi/2\le\theta<\pi/2$)
on the membrane viscosity $\eta_{\text {mb}}^*$ for reduced shear rate
$\dot\gamma^*=0.92$ and various reduced volumes $V^*$.
The error bars are estimated from three independent runs.
Squares and circles represent discocyte and prolate vesicles at 
$V^*=0.59$, respectively. Triangles and diamonds represent prolate vesicles 
at $V^*=0.78$ and $V^*=0.91$, where the prolate is the only stable shape.
The solid lines and broken line are calculated by K-S theory with 
prolate ($V^*=0.59, 0.78$, and $0.91$) and 
oblate ellipsoids ($V^*=0.59$), respectively.
}
\end{figure}

To induce a shear flow, we employed Lee-Edwards boundary condition 
\cite{kiku03,alle87}, which gives a linear flow profile 
$(v_x,v_y,v_z)=(\dot\gamma z,0, 0)$ in the MPCD fluid.
The particle density was set to $\rho=10m_{\text {s}}/a^3$ 
($N_{\text s}=450000$, $L_x=50a$, $L_y=L_z=30a$).
In experimental conditions of red blood cells and liposomes, 
the Reynolds number Re$=\dot\gamma\rho R_{\text 0}^2/\eta_{\text {0}}$ is 
very small (Re $\sim 10^{-3}$), where $R_{\text 0}=\sqrt{S/4\pi}$ is 
the effective vesicle radius. Therefore, 
we chose a short mean free path $h\sqrt{k_{\text B}T/m_{\text {s}}}=0.025 a$,
where $k_{\text B}$ is the Boltzmann constant and $T$ is the temperature 
\cite{ripo04}.
Then the viscosity of solvent fluid is 
$\eta_{\text {0}}= 20.1\sqrt{m_{\text {s}}k_{\text B}T}/a^2$ \cite{kiku03}.
We used $\kappa=20k_{\text B}T$, $N_{\text {mb}}=500$, and 
$m_{\text {mb}}=10m_{\text {s}}$. 
The volume $V$ and surface area $S=405 a^2$ of a vesicle are kept constant
to about 1\% accuracy. 
With these parameters, we obtain a Reynolds number Re$\simeq 0.1$.
The results are conveniently expressed in terms of dimensionless 
variables:
the reduced volume $V^*=V/(4\pi R_{\text 0}^3/3)$, the intrinsic time scale 
$\tau=\eta_{\text {0}}R_{\text 0}^3/\kappa$, 
the reduced shear rate $\dot\gamma^*=\dot\gamma\tau$, and the relative membrane 
viscosity $\eta_{\text {mb}}^*=\eta_{\text {mb}}/\eta_{\text 0}R_{\text 0}$. 
Details of the numerical scheme will be published elsewhere \cite{nogu04}.

At $\eta_{\text {mb}}^*=0$, a vesicle exhibits tank-treading motion for 
all simulated reduced volumes in the range $0.59\le V^*\le 0.97$. 
We calculated the average inclination angles $\langle\theta\rangle$, and found
them to agree very well with those obtained by the boundary integral method  
\cite{krau96}. 

With increasing membrane viscosity $\eta_{\text {mb}}^*$, the inclination 
angle $\theta$ decreases, as shown in Fig.~\ref{fig:ang}.
The qualitative features of the simulation data are reproduced very well
by the theory of Keller and Skalak 
(K-S) \cite{kell82,tran84}. Note that there are no adjustable parameters.
Due to the approximations in the K-S theory, an agreement on a 
quantitative level cannot be expected: (i) an ellipsoidal shape is 
assumed, which is only mimics the real shapes of vesicles, 
(ii) the flow on the surface of the
droplet is not locally area conserving, as it must be for a incompressible
membrane, and (iii) thermal fluctuations are ignored in the theory, but 
are present in the simulations.   
In the K-S theory, the vesicle transits from tank-treading to tumbling 
motion when the angle $\theta$ reaches $0$. In contrast, we observe 
tumbling intermittently to occur already for nonzero $\langle\theta\rangle$, 
since our simulation includes thermal fluctuation. 
For example, the vesicle with $V^*=0.78$ starts tumbling at 
$\eta_{\text {mb}}^*=1.22$.
This intermittent tumbling smoothes out the decrease in $\langle\theta\rangle$ 
around the transition point, compare Fig.~\ref{fig:ang}. 

\begin{figure}
\includegraphics{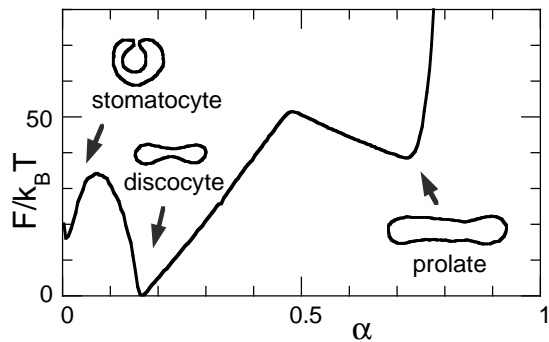}
\caption{ \label{fig:fe}
Free-energy profile  $F(\alpha)$ of the asphericity $\alpha$ for $V^* = 0.59$
in the absence of shear 
flow. The sliced snapshots of stable (discocyte) and metastable (prolate and 
stomatcyte) shapes are also shown. }
\end{figure}

We now focus on the case $V^*=0.59$. At this reduced volume, the oblate 
discocyte shape is stable and the prolate and 
stomatocyte shapes are metastable in the absence of shear flow. 
Fig.~\ref{fig:fe} shows the free energy $F$ as a function of 
the asphericity $\alpha$, calculated 
with a version of the generalized-ensemble Monte Carlo method \cite{berg03}. 
The asphericity 
$\alpha = [({\lambda_1}-{\lambda_2})^2+ ({\lambda_2}-{\lambda_3})^2+ 
({\lambda_3}-{\lambda_1})^2]/[2 R_{\text g}^4]$,
with the eigenvalues ${\lambda_1} \leq {\lambda_2} \leq {\lambda_3}$ of the 
moment-of-inertia tensor and the squared 
radius of gyration $R_{\text g}^2=\lambda_1+\lambda_2+\lambda_3$, is a
convenient measure to distinguish oblate and prolate shapes, where
$\alpha=0$ for spheres, $\alpha=1$ for thin rods, and $\alpha=0.25$ for  
thin discs \cite{rudn86}. The shear flow changes this stability.
For membrane viscosity $\eta_{\text {mb}}^*=0$ and shear rates 
$\dot\gamma^*\ge 1.66$, the discocyte state is found to be 
destabilized and to transform into a prolate,
in agreement with the results of Ref.~\cite{krau96}. However, for
a smaller shear rate of $\dot\gamma^*=0.92$, the discocyte vesicle 
retains its shape. Speculations about shear to be a singular 
perturbation \cite{krau96} can therefore be ruled out. 

\begin{figure}
\includegraphics{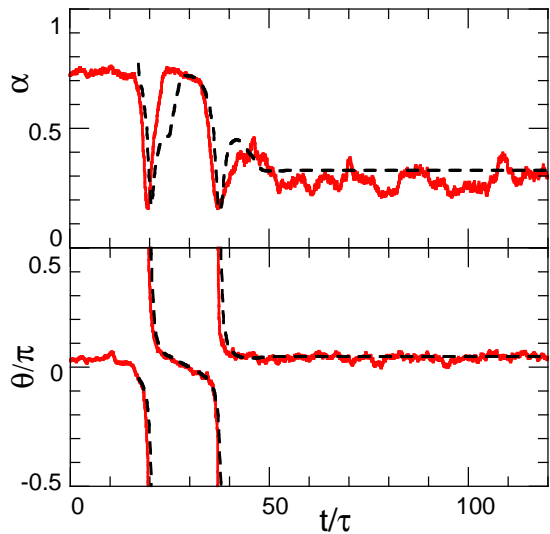}
\caption{\label{fig:d1}
(Color online)
Time development of (a) asphericity $\alpha$ and (b) inclination angle 
$\theta$, for $V^*=0.59$, $\eta_{\text {mb}}^*=1.62$ and $\dot\gamma^*=1.84$.
The broken lines are obtained from Eqs.~(\ref{eq:al}) and (\ref{eq:thet})
with $\zeta_{\alpha}=100$, $A=15$, and $B(\alpha)=1.1-0.18\alpha$.
 }
\end{figure}

\begin{figure}
\includegraphics{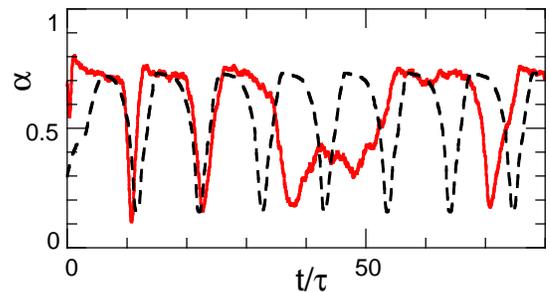}
\caption{\label{fig:d2}
(Color online)
Time development of (a) asphericity $\alpha$ and (b) inclination angle 
$\theta$ at $\dot\gamma^*=2.76$.
The other parameters are set to same values in Fig.~\ref{fig:d1}.
 }
\end{figure}

The inclination angle $\theta$ of prolates decreases faster than that 
of discocytes
with increasing $\eta_{\text {mb}}^*$, see Fig.~\ref{fig:ang}.
At a large membrane viscosity of $\eta_{\text {mb}}^*=1.62$, the 
prolate enters the tumbling phase,
while the discocyte remains in the tank-treading phase. 
The reason is that the discocyte has a flat dimple region and is 
less affected by the membrane viscosity than the prolate.
Remarkably, for small shear rates $\dot\gamma^*\le 1.84$, 
the (metastable) prolate starts tumbling, but after a $\pi$ or $2\pi$ 
rotation, transforms into a tank-treading discocyte, see Fig.~\ref{fig:d1}.
Only for larger shear rates of $\dot\gamma^*\ge 2.76$, the tumbling 
continues --- accompanied 
by shape oscillations between prolate and discocyte (Fig.~\ref{fig:d2}).
At intermediate membranes viscosities, $\eta_{\text {mb}}^*=0.49$ and
$\eta_{\text {mb}}^*=0.87$, and shear rate $\dot\gamma^*=0.92$, the 
prolate transforms into a discocyte 
after tank-treading motion for a time of  
$(70\pm40)\tau$ or $(40\pm20)\tau$ by thermal fluctuation, respectively.
For a larger shear rate of $\dot\gamma^*=1.84$, the tumbling continues 
intermittently.

K-S theory~\cite{kell82,tran84} explains the 
$\eta_{\text {mb}}^*$-dependence of the stability of tank-treading 
(compare Fig.~\ref{fig:ang}), but cannot be applied to describe 
the dynamics, including morphological changes. 
We suggest the simple equations
\begin{eqnarray}
\zeta_{\alpha} \dot{\alpha}= 
-\kappa^{-1}\partial F/\partial \alpha + A\dot\gamma^* \sin(2\theta) 
\label{eq:al}\\
\dot{\theta}=0.5\dot\gamma^*\{-1+B(\alpha) \cos(2\theta)\} \label{eq:thet}.
\end{eqnarray}
The force $\partial F/\partial \alpha$ is calculated from the free 
energy $F(\alpha)$ of Fig.~\ref{fig:fe}.
The second term of Eq.~(\ref{eq:al}) is the deformation force due to the
shear flow. Its angular dependence can be derived from the shape 
equations of Ref.~\cite{krau96}, while the amplitude is assumed to 
be independent of the asphericity $\alpha$ (to leading order). 
Eq.~(\ref{eq:thet}) is adopted from K-S theory \cite{kell82,tran84}. 
Here, $B$ is a constant which depends on viscosities and ellipsoid shape.
For $B>1$, a steady angle $\theta=0.5\arccos(1/B)$ exists and 
tank-treading motion occurs, while 
for $B<1$, there is no stable angle and tumbling motion occurs.
In our case, the vesicle shape can be time dependent, so that $B$ is
no longer constant. For simplicity, we assume a 
linear dependence of $B$ on the asphericity, $B(\alpha)=B_0-B_1\alpha$.
To obtain tank-treading discocytes and tumbling prolate, we need
$B(0.2)>1$ and $B(0.7)<1$, respectively.
Then, Eqs.~(\ref{eq:al}) and (\ref{eq:thet}) reproduce the simulated 
dynamics very well, see Figs.~\ref{fig:d1} and \ref{fig:d2}.
The vesicle is found, for example, to relax after some tumbling to a stable, 
tank-treading discocyte state at $\dot\gamma^*=1.84$, and to  
relax to a limit-cycle oscillation at $\dot\gamma^*=2.76$.

The vesicle deformation due to shear depends on the inclination 
angle $\theta$. Shear flow increases the elongation of a vesicle for 
$0<\theta<\pi/2$ (where $\dot\gamma^*\sin(2\theta)>0$),
and the angles $\theta$ of tank-treading motion belong to this region.
On the other hand, shear flow reduces the elongation for $-\pi/2<\theta<0$ 
(where $\dot\gamma^*\sin(2\theta)<0$) during tumbling.
The force in the former case induces the discocyte-to-prolate transformation,
in the latter case the prolate-to-discocyte transformation.
With increasing membrane viscosity $\eta_{\text {mb}}^*$, 
the inclination angle $\theta$ of a tank-treading discocyte decreases,
and larger shear rates $\dot\gamma^*$ are necessary to generate the required 
elongational forces to induce a discocyte-to-prolate transition.

It is also interesting to compare the effect of membrane viscosity 
$\eta_{\text {mb}}$ and internal viscosity $\eta_{\text {in}}$.
In both cases, an increase of the viscosity 
induces a decrease of the inclination angle $\theta$ and a transition from
tank-treading to tumbling.
However, the effect of internal viscosity $\eta_{\text {in}}$ is less 
dependent on the vesicle morphology.
K-S theory \cite{kell82} shows that
the tank-treading phase of oblate vesicle is destabilized a little 
faster than that of prolate
with an increase of $\eta_{\text {in}}$ (for $\eta_{\text {mb}}^*=0$);
the transition viscosity is $\eta_{\text {in}}/\eta_{\text {0}}= 2.8$ 
and $3.3$ of oblate and prolate ellipsoids at $V^*=0.59$, respectively.
Thus, only a sufficiently high membrane viscosity $\eta_{\text {mb}}^*$ 
can induce a prolate-to-discocyte transformation.

The membrane viscosity of human red blood cells is estimated from  
the analysis of the tank-treading motion to be 
$\eta_{\text {mb}}=10^{-7}$Ns/m \cite{tran84}, while a  
micropipette recovery-time technique gives 
$\eta_{\text {mb}}=10^{-6}$Ns/m \cite{nash83}.
When the viscosity of the external fluid is set to the same value of
the intracellular fluid,
$\eta_{\text 0}=10^{-2}$Pa~s, and $R_{\text 0}=3.3\mu$m,
the relative membrane viscosity is then found to be in the range 
$\eta_{\text {mb}}^*=1...10$.
Thus, the effect of this membrane viscosity is sufficiently large (compare 
Fig.~\ref{fig:ang}) to strongly affect the dynamics of erythrocytes.
The viscoelasticity of membrane can be changed by varying the chemical 
composition of the solvent \cite{chie87}.
It is difficult to separate the effects of viscosity and elasticity, however.
On the other hand, the membrane viscosity of polymersome can be changed 
by varying the polymer chain length. Thus,  
polymersomes seem to be very well suited to study the effect of 
membrane viscosity experimentally.

To summarize, we have applied the MPCD method to a vesicle with viscous 
membrane under a simple shear flow.
The membrane viscosity qualitatively changes the vesicle dynamics.
The shear induces a discocyte-to-prolate or prolate-to-discocyte
transformation at low or high membrane viscosity, respectively.
The deformations of other vesicles shapes, such as stomatocyte and budded 
vesicles, will be interesting subjects in the further studies.

\begin{acknowledgments}
We would like to thank N.~Kikuchi (Oxford), A.~Lamura (Bari), and R.G. Winkler 
for helpful discussions. 
HN's stay at FZJ was supported by the Postdoctoral Fellowships for Research 
Abroad of the Japan Society for the Promotion of Science (JSPS).
\end{acknowledgments}

\end{document}